\documentclass[aps,pra,preprint]{revtex4-2}
\usepackage{graphicx}
\usepackage{color}
\usepackage{amsmath,amssymb}
\begin{document}

\title{Efficient transfer of entanglement along a qubit chain in the presence of thermal fluctuations}
\author{Gourab Das}
\author{Rangeet Bhattacharyya}
\email{rangeet@iiserkol.ac.in}
\affiliation{Department of Physical Sciences, Indian Institute of Science Education and Research Kolkata,\\
Mohanpur -- 741246, WB, India}

\begin{abstract}

Quantum communications require efficient implementations of quantum state transportation with high fidelity.
Here, we consider the transport of entanglement along a chain of qubits. A series of SWAP operations
involving successive pairs of qubits can transport entanglement along the chain. We report that the fidelity
of the abovementioned gate has a maximum value corresponding to an optimum value of the drive amplitude in
the presence of drive-induced decoherence. To incorporate environmental effect, we use a previously reported
fluctuation-regulated quantum master equation [A. Chakrabarti and R. Bhattacharyya, Phys. Rev. A 97, 063837
(2018)]. The existence of an optimum drive amplitude implies that these series of SWAP operations on open
quantum systems would have an optimal transfer speed of the entanglement. 

\end{abstract}

\maketitle

\section{Introduction}

Transfer of coherences and entanglements along a qubit-network is one of the important aspects of the
quantum information processing. 
%A qubit network could be modeled as a dipolar-coupled ising spin-chain of spin-$\frac{1}{2}$ particles.
We investigate the efficiency of the transfer of coherences and entanglements along the spin-chain in a
dissipative environment. To this end, we use a fluctuation-regulated quantum master equation (frQME) to
include the higher order effects of the drive and the dipolar interactions. We show that for the coherence
transfer there exists a set of optimal conditions on the drive parameters. For the entanglement transfer,
the efficiency of the transfer shows the existence of an optimal condition involving the drive amplitude and
the system-bath coupling. This transport problem becomes trivial in presence of a SWAP-gate. In this work we
tried to simulate the mentioned gate in presence of a dissipative environment.

Sometimes dipolar interaction among our Rydberg system prevents flow of information of excitement.  This
phenomenon is known as Rydberg blockade. Blockade mechanism relies on the strong interaction between
different parts of a system. Blockade has been shown in cold atoms\cite{Schlosser_2001},
photons\cite{Birnbaum_2005} and electrons\cite{Fulton_1987},\cite{Averin_1986},\cite{Ono_2002}.
\cite{Urban_2009} has demonstrated a single neutral rubidium atom (Rydberg excited) blocks excitation of a
second atom that is located more than 10 µm away. They have observed probability of double excitation going
down below 20$\%$. Also \cite{Gaetan_2009} has demonstrated Rydberg blockade between two atoms trapped atoms
(by optical tweezers) at 4 µm  distance. They have shown experimentally that two-atom’s collective behavior
enhances single-atom excitation.

Fast two-qubit quantum gate using neutral atoms has been implemented \cite{Jaksch_2000}. Their gate
operation time is much shorter than the external motion of the atoms in trapping potential. Their
requirement of large interaction energy is provided by dipole-dipole interaction between Rydberg atoms.
Optically excited strong dipolar coupled atoms results dipolar blockade. This has been used for coherent and
entanglement manipulation for cold Rydberg atoms \cite{Lukin_2001}.

Ultracold Rydberg atoms can produce blockade effect \cite{Schlosser_2001}. The opposite effect of blockade
is known as anti-blockade effect. Antiblockade has been demonstrated using Autler-Townes double peak
structure \cite{Ates_2007}. They have used an optical lattice created by CO2 lasers for antiblockade. Also
laser excited three level Rydberg gas shows antiblockade \cite{Amthor_2010}. One dimensional array of
Rydberg atom with van der Waals type interaction also shows blockade and antiblockade under periodic
modulation \cite{Basak_2018}.

Quantum state has been transferred on spin-$\frac{1}{2}$ one dimensional chain with Heisenberg coupling
\cite{Banchi_2011}. They have shown that the interaction strength controls the amount of transfer, for
low strength there is only Rabi like oscillations, for intermediate coupling there is no transfer and strong
coupling shows intrachannel transfer. Also the amount of entanglement per block of certain length on a long
spin 1 Heisenberg chain has been calculated in thermodynamic limit \cite{Orus_2008}.

We discussed frQME in section 2. Our model system, SWAP operation and its fidelity optimization are discussed
in section 3. Then we discuss our results in section 4.

\section{frQME }
\label{sec:4}

We present a brief derivation of FRQME \cite{Chakrabarti_2018}.  In open quantum systems, one assumes the
thermal fluctuation happens in a thermal reservoir and can be represented by a suitable chosen local
environment Hamiltonian that represents a driven-dissipative system. Hence,  the general form of the
Hamiltonian for the system and the local environment (in frequency unit) can be represented as

\begin{equation}
\mathcal{H}(t) = \mathcal{H}_{\text{S}}^0 +  \mathcal{H}_{\text{E}}^0 +  \mathcal{H}_{\text{SE}} + 
\mathcal{H}_{\text{S}}(t) + \mathcal{H}_{\text{E}}(t) \label{4.1}
\end{equation}

where $\mathcal{H}_{\text{S}}^0$ is the time-independent Hamiltonian of the system,
$\mathcal{H}_{\text{E}}^0$ is the time-independent Hamiltonian of the local environment, $
\mathcal{H}_{\text{SE}}$ is the coupling Hamiltonian between the system with its local environment having
strength $\omega_{\text{SE}}$,  $\mathcal{H}_{\text{S}}(t)$ is the external drive Hamiltonian applied on the
system with amplitude $\omega_1$ and $\mathcal{H}_{\text{E}}(t)$ is the fluctuations Hamiltonian for local
environment. $\mathcal{H}_{\text{E}}(t)$ is chosen to be in eigenbasis ${| \phi_j \rangle}$ of
$\mathcal{H}_{\text{E}}^0$, as it is assumed that the fluctuations should not drive the lattice away from
equilibrium.  Hence,

\begin{equation}
\mathcal{H}_{\text{E}}(t) = \sum_j f_j (t) |\phi_j\rangle \langle \phi_j | \label{4.2}
\end{equation}

where $f_j (t)$'s are independent, Gaussian, $\delta$-correlated stochastic variables with zero mean and
standard deviation $\kappa$, i.e. $\overline{f_j (t)} = 0$ and $\overline{f_j (t_1) f_j (t)} = \kappa^2
\delta(t_1 - t_2)$. 

If we consider the whole dynamics of our system with its environment, we can write the evolution using
Liouville-von Neumann equation,
\begin{equation}
\frac{d}{dt} \tilde{\rho} (t) = - i [H (t), \tilde{\rho}(t)]\label{4.3} 
\end{equation}
where $\tilde{\rho}(t)$ is the density matrix of system-environment combined in interaction representation
and $H (t)$ is the total Hamiltonian in interaction picture too.  We take trace over environment variables
to get dynamics (system density matrix) of the system from time $t$ to $t+\Delta t$,
\begin{equation}
\tilde{\rho}_S (t+\Delta t) = \tilde{\rho}_S (t) - i \int_t^{t+\Delta t} dt_1 Tr_E [H_{\rm eff} (t_1),\tilde{\rho} (t)]\label{4.4}
\end{equation}
where, $H_{\rm eff} (t)$ is the interaction picture representation of $\mathcal{H}_{\rm eff} (t) =
\mathcal{H}_S (t) + \mathcal{H}_{SE} (t)$.  Partial tracing over environmental degrees of freedom kills the
commutator involving $H_E (t_1)$.  Using the time evolution operator of the system-environment duo, we can
write the density matrix at time $t_1$ as $\tilde{\rho} (t_1) = U (t_1,t) \tilde{\rho}(t)
U^{\dagger}(t_1,t)$.  We simplify our notation by using $U(t_1)$ instead of $U(t_1,t)$.  Hence, from
Sch{\"o}dinder equation we can write,
\begin{eqnarray}
U(t_1) &=& \mathcal{I} - i \int_t^{t_1} H(t_2) U(t_2) dt_2 \nonumber\\
&=& \mathcal{I} - i \int_t^{t_1} H_{\rm eff}(t_2) U(t_2) dt_2 - i \int_t^{t_1} H_{E}(t_2) U(t_2) dt_2\label{4.5}
\end{eqnarray}
One assumes $t_1 - t << \frac{1}{\omega_1},\frac{1}{\omega_{SE}}$ and neglects the effect of $H_{\rm eff}$
in the time interval from $t$ to $t_2$; i.e. we can approximate $U(t_2)$ by $U_E (t_2)$. Hence we can write
\eqref{4.5} as,
\begin{eqnarray}
U(t_1) &\approx & \mathcal{I} - i \int_t^{t_1} H_{\rm eff}(t_2) U_E(t_2) dt_2 - i \int_t^{t_1} H_{E}(t_2) U_E(t_2) dt_2  \nonumber\\
&=& U_E(t_1) - i \int_t^{t_1} H_{\rm eff}(t_2) U_E(t_2) dt_2 \label{4.6}
\end{eqnarray}
where we used $U_E(t_1) =  \mathcal{I} - i \int_t^{t_1} H_{E}(t_2) U_E(t_2) dt_2$.

We substitute \eqref{4.6} in \eqref{4.4} to get,
\begin{eqnarray}
\tilde{\rho}_S (t+\Delta t)&=& \tilde{\rho}_S (t) - i \int_t^{t+\Delta t} dt_1 Tr_E [H_{\rm eff}(t_1), U_E(t_1) \tilde{\rho} (t) U_E^{\dagger}(t_1)] \nonumber\\
&& - \int_t^{t+\Delta t} dt_1 \int_t^{t_1} dt_2 Tr_E [H_{\rm eff} (t_1), H_{\rm eff} (t_2) U_E(t_2) \tilde{\rho}(t)U_E^{\dagger}(t_1) - U_E(t_1) \tilde{\rho}(t)U_E^{\dagger}(t_2) H_{\rm eff} (t_2)] \nonumber\\ &&+ O(H^3_{\rm eff})\label{4.7}
\end{eqnarray}
where we ignored the terms with cubic or higher powers of $H_{\rm eff}$, denoted by $O(H^3_{\rm eff})$.
Using Born approximation the density matrix of system-environment duo can be factorized in its subsystems,
i.e.
\begin{equation}
\rho (t) = \rho_S (t) \otimes \rho_E^{eq}\label{4.8}
\end{equation}
where $\rho(t) =  \overline{\tilde{\rho}(t)}$. The Born approximation with the nature of the environmental
fluctuation provide the regulator in the second order under an ensemble average as
\begin{equation}
\overline{U_E(t_1) \tilde{\rho}(t) U_E^{\dagger}(t_2)} = \rho_S(t)\otimes \rho_E^{eq} e^{-\kappa^2 |t_1 - t_2|/2} \label{4.9}
\end{equation}
We perform coarse-graining procedure \cite{Cohen_1993} after substituting \eqref{4.9} in \eqref{4.7} to get FRQME,
\begin{eqnarray}
\frac{d}{dt} \rho_S (t) = &-& Tr_E [H_{\rm eff} (t), \rho_S (t) \otimes \rho_E^{eq}]^{sec} \nonumber\\
&-& \int_0^{\infty} d\tau Tr_E [H_{\rm eff}(t),[H_{\rm eff}(t-\tau), \rho_S (t) \otimes \rho_E^{eq}]]^{sec} e^{-|\tau | / \tau_c} \label{4.10}
\end{eqnarray}
where $\tau_c = 2/\kappa^2$.  Here we have ignored the fast oscillating terms as part of secular
approximation, denoted by the superscript "sec".  As $H_{\rm eff}$ contains drive term, it produces
drive-induced-dissipation (DID) in second order.  Also \eqref{4.10} is in GKLS form hence it's trace
preserving and completely positive. Just to mention, time-scale separation is also assumed, i.e. $\omega_1
\tau_c << 1$ and $\omega_{SE} \tau_c << 1$.

\section{Description of the system and SWAP operation}
\label{sec:7}

Our problem of finding optimal drive for transferring coherence or entanglement on spin-$\frac{1}{2}$ chain
become trivial if we can find drive for SWAP gate.  In this work we are considering spin chain with dipolar
coupling in NMR,  for which the the rotating wave approximated Hamiltonian depends on the Larmor
frequencies. Hence we will divide the implementation in two regimes, i.e. when the Larmor frequencies are
nearly same and they are far off, namely on- and far-resonance respectively.  These two regimes are
determined by inverse of coarse-grained time scale, $\Delta t$.  Not to forget due to environmental
fluctuation we will loose some information during this process.

\subsection{For two qubits having different Larmor frequency}\label{subsec:7.2}

First we discuss the case when Larmor frequencies of the spin-halves are far off.  Hence the secular
approximation to first order Liouvillian leads to dropping first quantum terms alongside second quantum
ones, i.e. it reduces to Heisenberg coupling Hamiltonian as if spin-chain is lying along $z$-axis,  i.e.
$\hat{\mathcal{H}}_J =  2\pi J \cdot I_z \otimes I_z \label{eq:7.2}$. The pulse sequence for a Heisenberg
coupled spin chain was known \cite{Madi_1998}
and corresponding unitary operator for the sequence is 

$\mathit{\tilde{U}} = \begin{pmatrix}
 -i & 0 & 0 & 0\\
 0 & 0 & i & 0\\
 0 & i & 0 & 0\\
 0 & 0 & 0 & i\\
\end{pmatrix} 
$. 

\begin{figure}[h!]
	\centering
	\includegraphics[width=0.9\textwidth]{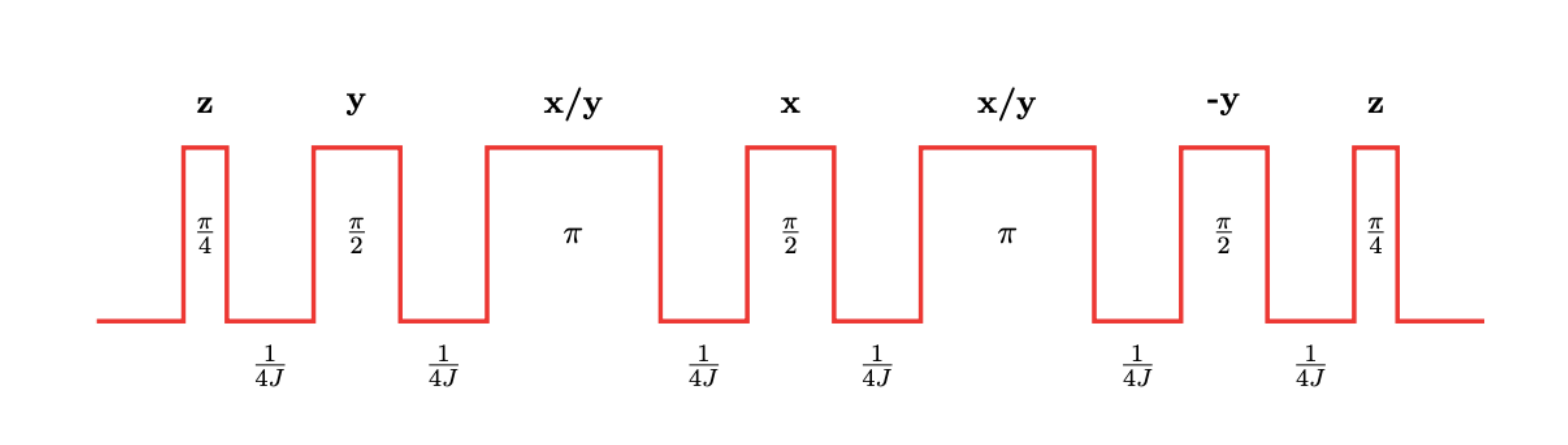}
	\hspace{1mm}
	\caption{Modified pulse sequence of SWAP gate  for non-identical qubits with dipolar coupling} 
	\label{fig:SWAP}
\end{figure}

We modified the sequence by two $\frac{\pi}{4}$ pulses along $z$-direction at two ends,  as shown in Fig.
\ref{fig:SWAP}. We get unitary operator for corresponding pulse sequence as, $ \mathit{U} = e^{- i
\frac{\pi}{4}}\cdot \mathit{U}_{swap}$, where 

$ \mathit{U}_{swap} = \begin{pmatrix}
 1 & 0 & 0 & 0\\
 0 & 0 & 1 & 0\\
 0 & 1 & 0 & 0\\
 0 & 0 & 0 & 1\\
\end{pmatrix} $ and the pre-factor represents a global phase factor.

Now consider a chain of three dipolar coupled spin halves. Imagine initially $|\psi_-\rangle =
\frac{1}{\sqrt{2}} (|10\rangle - |01\rangle)$ is shared between first and second spin and third spin is in
$|0\rangle$ state,  i.e. initial state $|\psi_i\rangle = \frac{1}{\sqrt{2}} (|10\rangle - |01\rangle)
\otimes |0\rangle$. We are trying to send $|\psi_-\rangle$ state between second and third spin,  i.e.  final
state $|\psi_f\rangle = |0\rangle \otimes \frac{1}{\sqrt{2}} (|01\rangle - |10\rangle)$.  Hence we can use
SWAP gate between first and third spin for the job.  So, effectively we have transported $|\psi_-\rangle$
state to second and third spins.  Here the effects of nearest neighbour couplings can be neutralized by a
short-lived $\pi$-pulse at the mid point of the evolution along $x$- or $y$-direction on second spin
\cite{bax_1984}.

\begin{figure}[h!]
%\subfloat[Fidelity of $|\psi_f\rangle$]% on $\omega_1$ and $\tau_c$
\raisebox{3cm}{(a)}\hspace{-1mm}
{\includegraphics[width=0.4\textwidth, keepaspectratio]{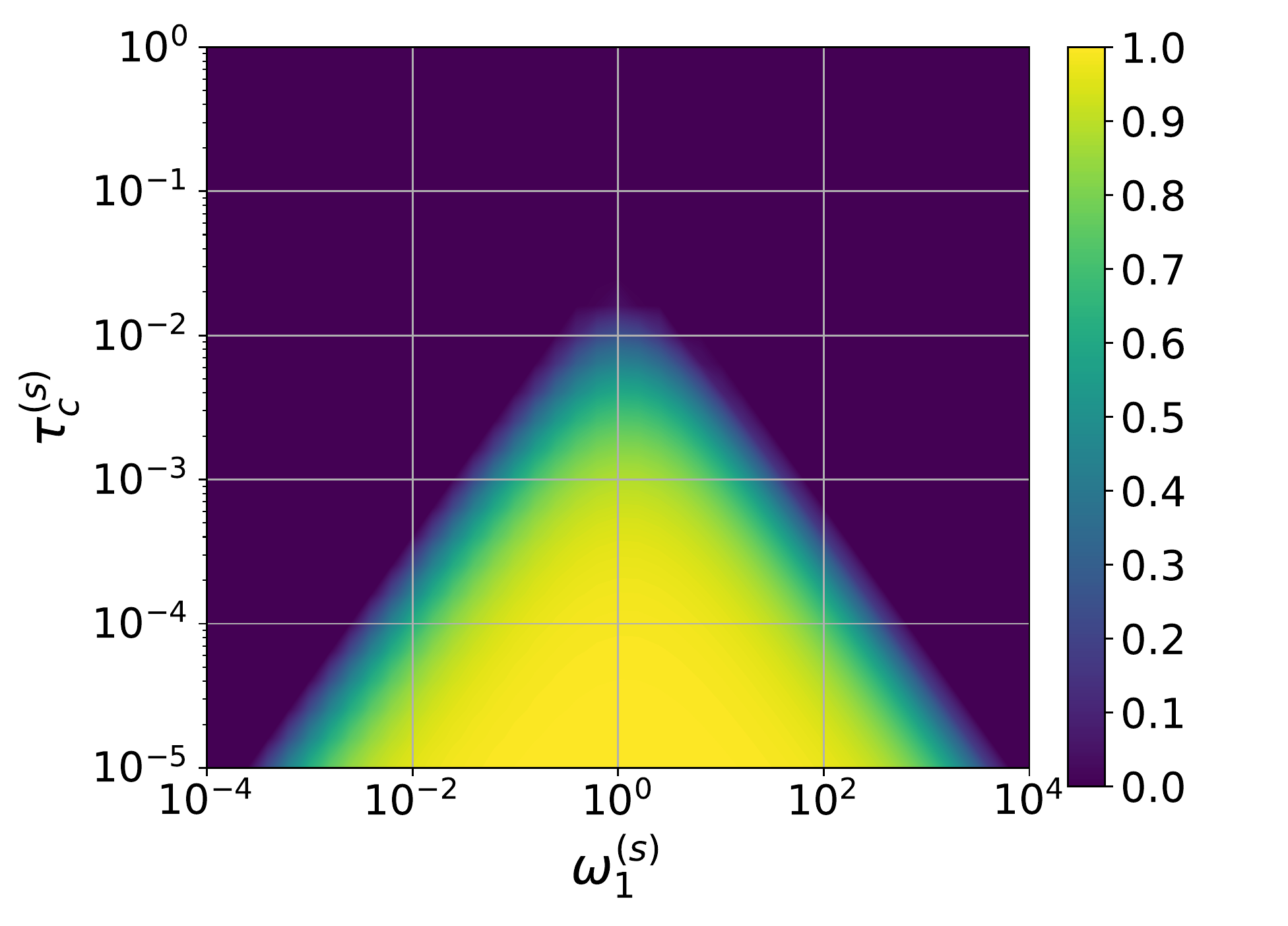}\label{fig:Fidelity}}
%\subfloat[Concurrence between 2nd \& 3rd spins] % on $\omega_1$ and $\tau_c$]\
\raisebox{3cm}{(b)}\hspace{-1mm}
{\includegraphics[width=0.4\textwidth, keepaspectratio]{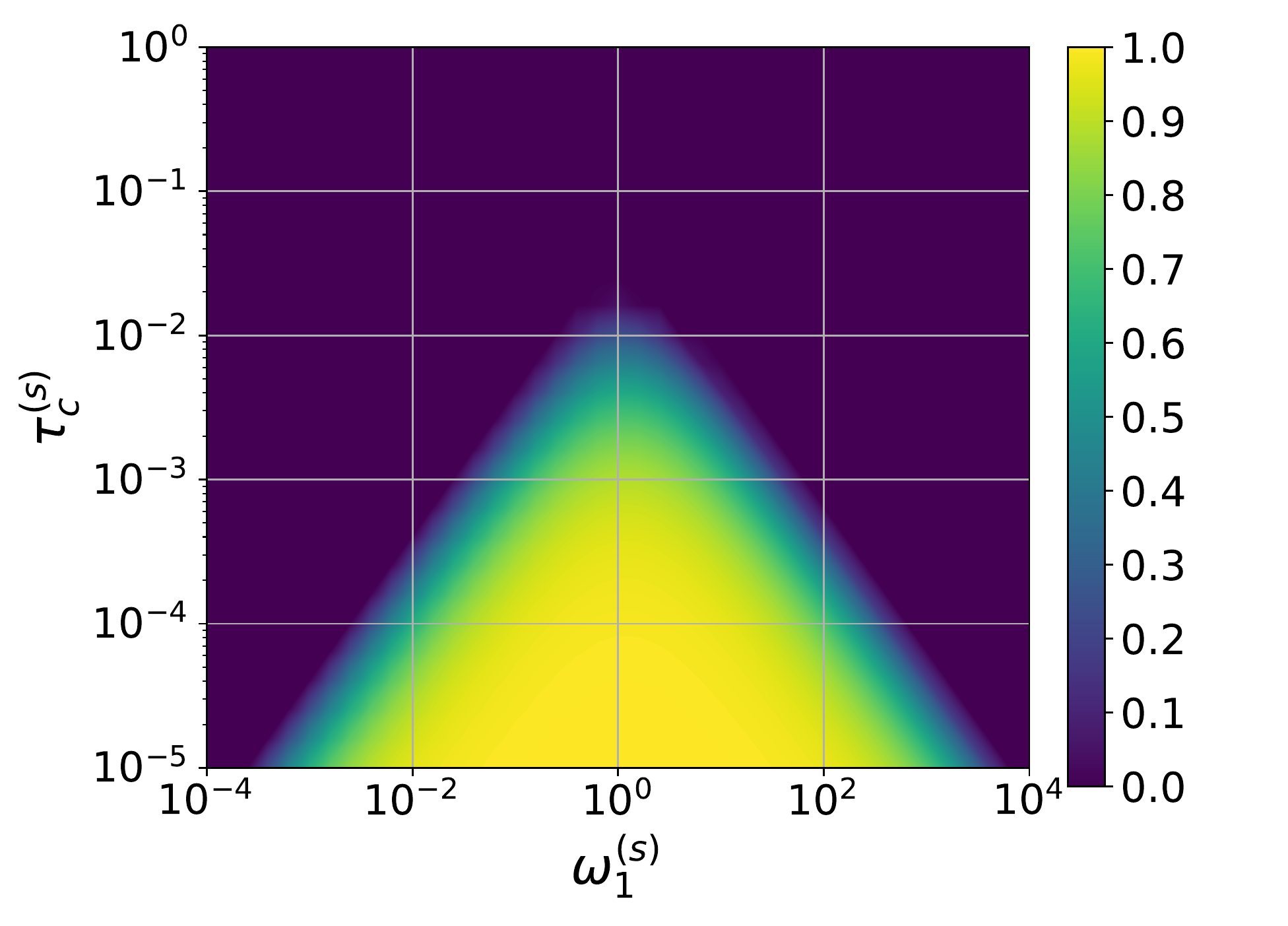}\label{fig:Concurrence}}\\
%\subfloat[Efficiency of SWAP gate]% on $\omega_1$ and $\tau_c$
\raisebox{3cm}{(c)}\hspace{-1mm}
{\includegraphics[width=0.4\textwidth, keepaspectratio]{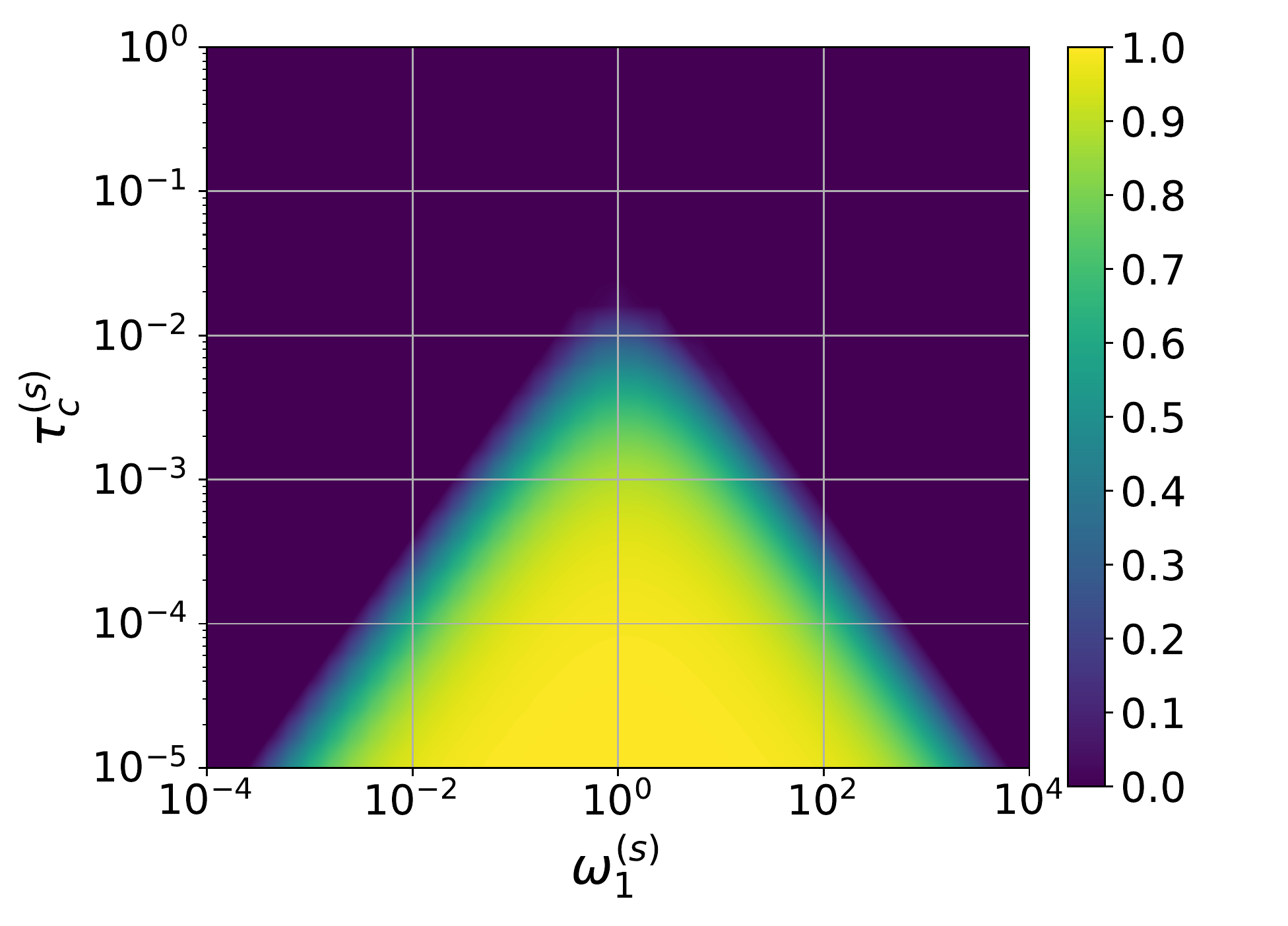}\label{fig:Efficiency}}
%\hspace{\fill}
%\subfloat[Fidelity of $|\psi_f\rangle$]% on $\omega_1$ and $\tau_c$
\raisebox{3cm}{(d)}\hspace{-1mm}
{\includegraphics[width=0.4\textwidth, keepaspectratio]{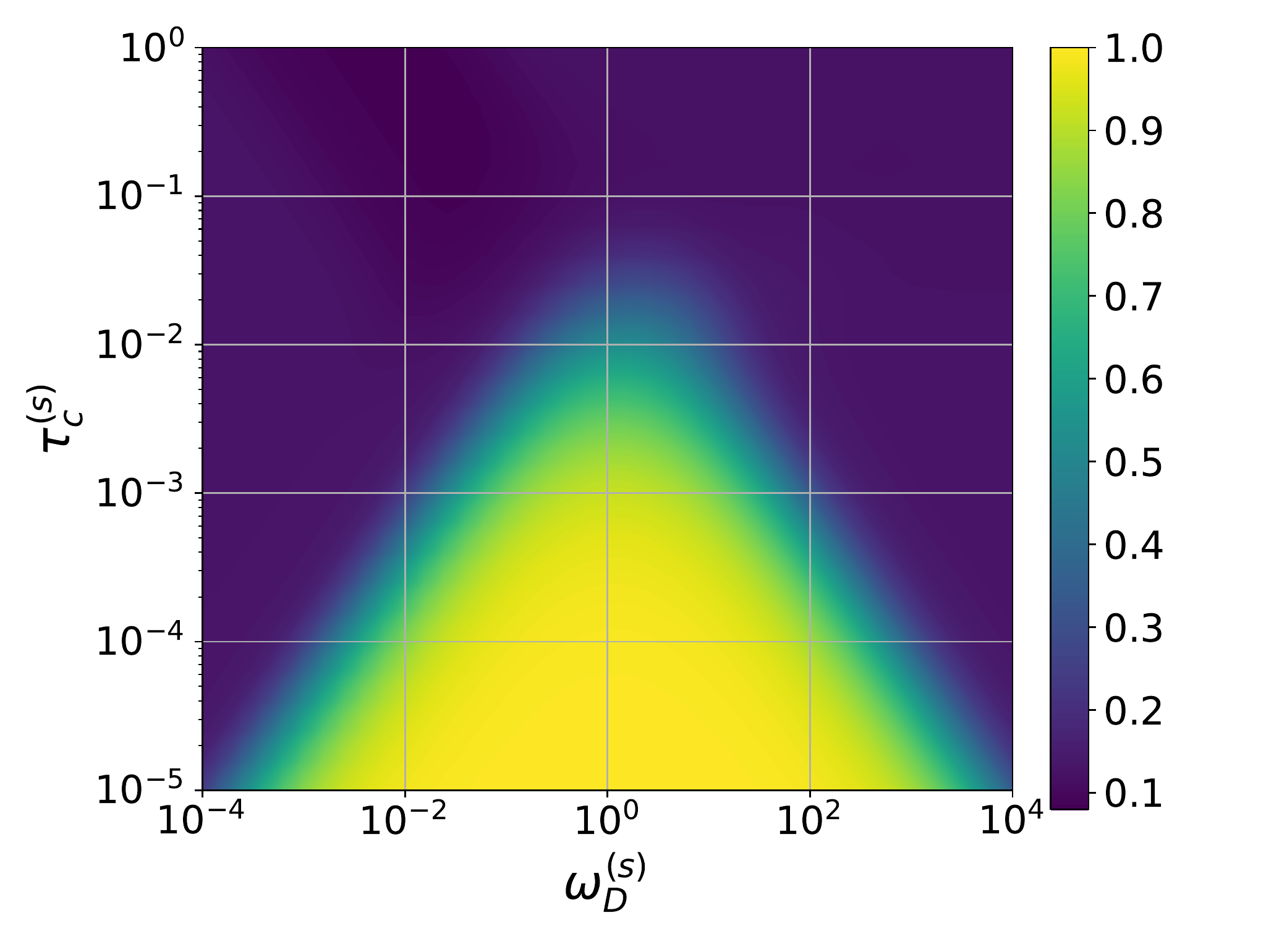}\label{fig:WJ_Fidelity}}

\caption{Concurrence between 2nd \& 3rd spin, Fidelity of expected state and efficiency as a function of
drive power ($\omega_1$) and characteristic time ($\tau_c$) of the bath (scaled to system-environment
coupling strength $\omega_{SE} $).  (a) Fidelity of $|\psi_f\rangle$, (b) Concurrence between 2nd \& 3rd
spins, (c) Efficiency of SWAP gate, (d) Fidelity of $|\psi_f\rangle$.  These simulated results are for
Larmor frequencies of the spins as $\omega_0^1 = 2\pi \times 10^4$ kHz, $\omega_0^2 = 2\pi \times 10^3$ kHz,
$\omega_0^3 = 2\pi \times 5 \times 10^2$ kHz, strength of system-environment coupling $\omega_{SE} = 2\pi
\times 10^2$ kHz.  (a), (b) \& (c) are for
$\omega_D = 2\pi \times 150$ kHz and and (d) is for $\omega_1 = 2\pi \times 150$
kHz.}

\label{fig:Hei_SWAP}
\end{figure}

Fig.\ref{fig:Hei_SWAP} shows we can get very high efficiency as well as fidelity of expected state (and
hence high entanglement transfer) when drive's power ($\omega_1$) and dipolar coupling strength ($\omega_D$)
are similar to system-environment coupling strength ($\omega_{SE}$) and for certain characteristic time of
bath ($\tau_c$) for this case too.  Also environmental effects induces additional decoherence.

\subsection{For two qubits having the same Larmor frequency}\label{subsec:7.1}

\begin{figure}[h!]
	\centering
	\includegraphics[width=.9\textwidth]{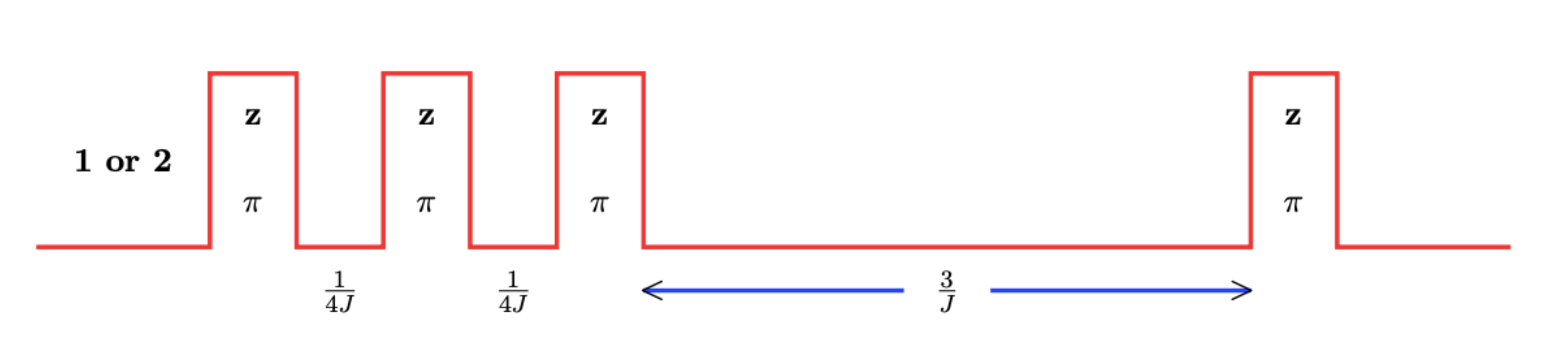}
	\hspace{1mm}
	\caption{Pulse sequence of SWAP gate for identical qubits with dipolar coupling} 
	\label{fig:DI_SWAP}
\end{figure}

Now we consider the case when the Larmor frequencies of the spin-halves,dipolar coupled, are same.  Hence if
the chain is lying along $x$-axis, so the coupling Hamiltonian is  $\hat{\mathcal{H}}_D =  2\pi J \cdot (I_z
\otimes I_z - \frac{1}{4} (I_+ \otimes I_- + I_- \otimes I_+) -  \frac{3}{4} (I_+ \otimes I_+ + I_- \otimes
I_-))\label{eq:7.3}$.  To construct the pulse sequence in Fig.  \ref{fig:DI_SWAP} we have considered
zero-quantum terms of the coupling as secular approximation to first order Liouvillian, for this case which
corresponds to dropping second quantum terms from the above Hamiltonian.  In this case, the unitary operator
is  $\mathit{U} = e^{- i \frac{3\pi}{4}}\cdot \mathit{U}_{swap}$. Note, here we need to drive just one of
the coupled spins. 

Now consider the similar situation as discussed in subsection \ref{subsec:7.1} but the Larmor frequencies
are far off, i.e.  initially $|\psi_i\rangle = \frac{1}{\sqrt{2}} (|10\rangle - |01\rangle) \otimes
|0\rangle$ is shared between three dipolar coupled spin-halves and we apply pulses on first and third spins
corresponding to SWAP gate between them to get the final state $|\psi_f\rangle = |0\rangle \otimes
\frac{1}{\sqrt{2}} (|01\rangle - |10\rangle)$.  So, effectively we have transported $|\psi_-\rangle$ state
to second and third spins.  Here also one can neutralize nearest neighbour couplings' effect by short-lived
$\pi$-pulse(s) along $z$-direction on second spin (as earlier). Here we have used frQME
\cite{Chakrabarti_2018} to incorporate environment effect.

\begin{figure}[h!]   
%\subfloat[Fidelity of $|\psi_f\rangle$]% on $\omega_1$ and $\tau_c$
\raisebox{3cm}{(a)}\hspace{-1mm}
\includegraphics[width=0.4\textwidth, keepaspectratio]{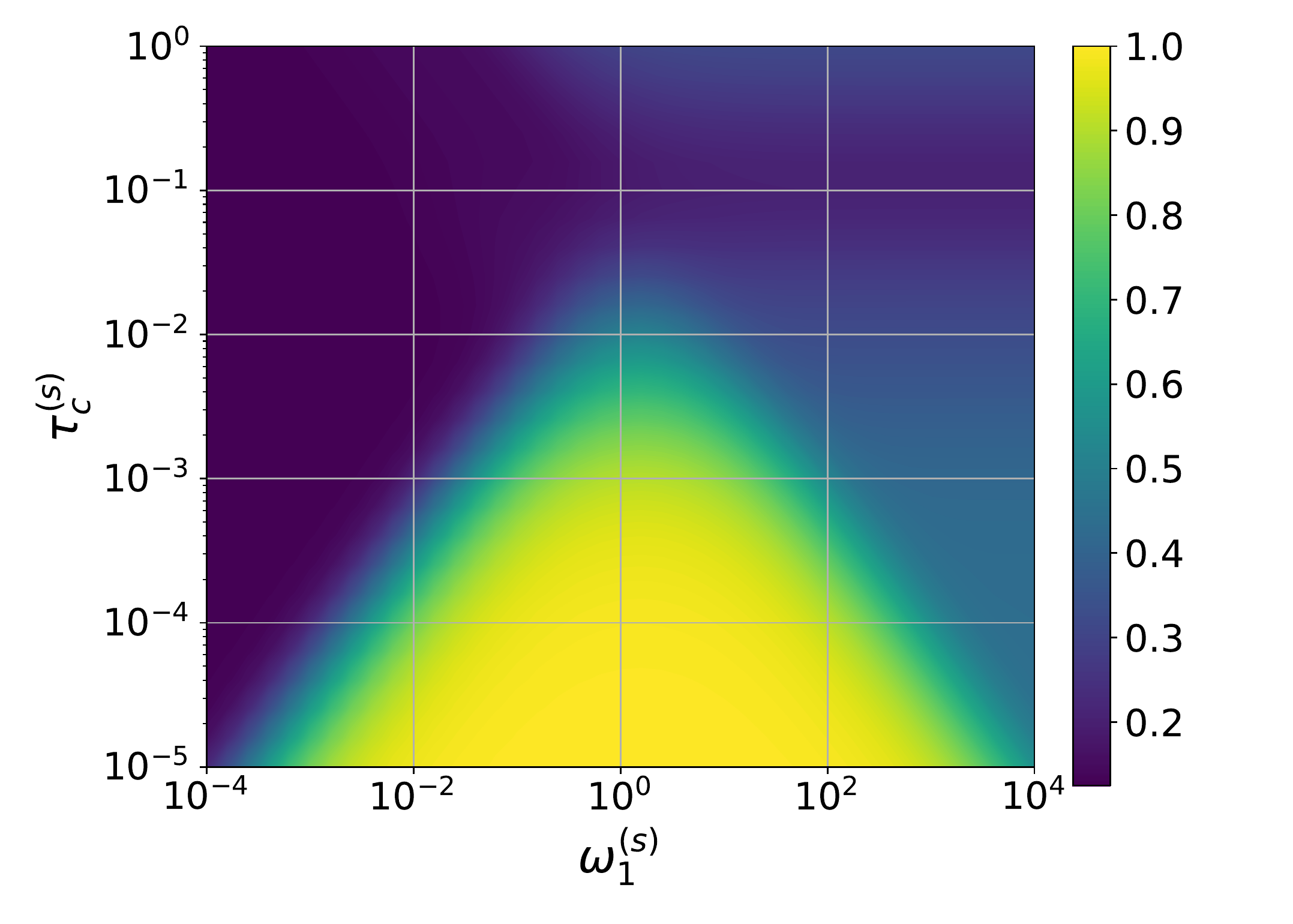}%\label{fig:DI_Fidelity}}
%\hspace{\fill}
%\subfloat[Concurrence between 2nd \& 3rd spin]% on $\omega_1$ and $\tau_c$
\raisebox{3cm}{(b)}\hspace{-1mm}
\includegraphics[width=0.4\textwidth, keepaspectratio]{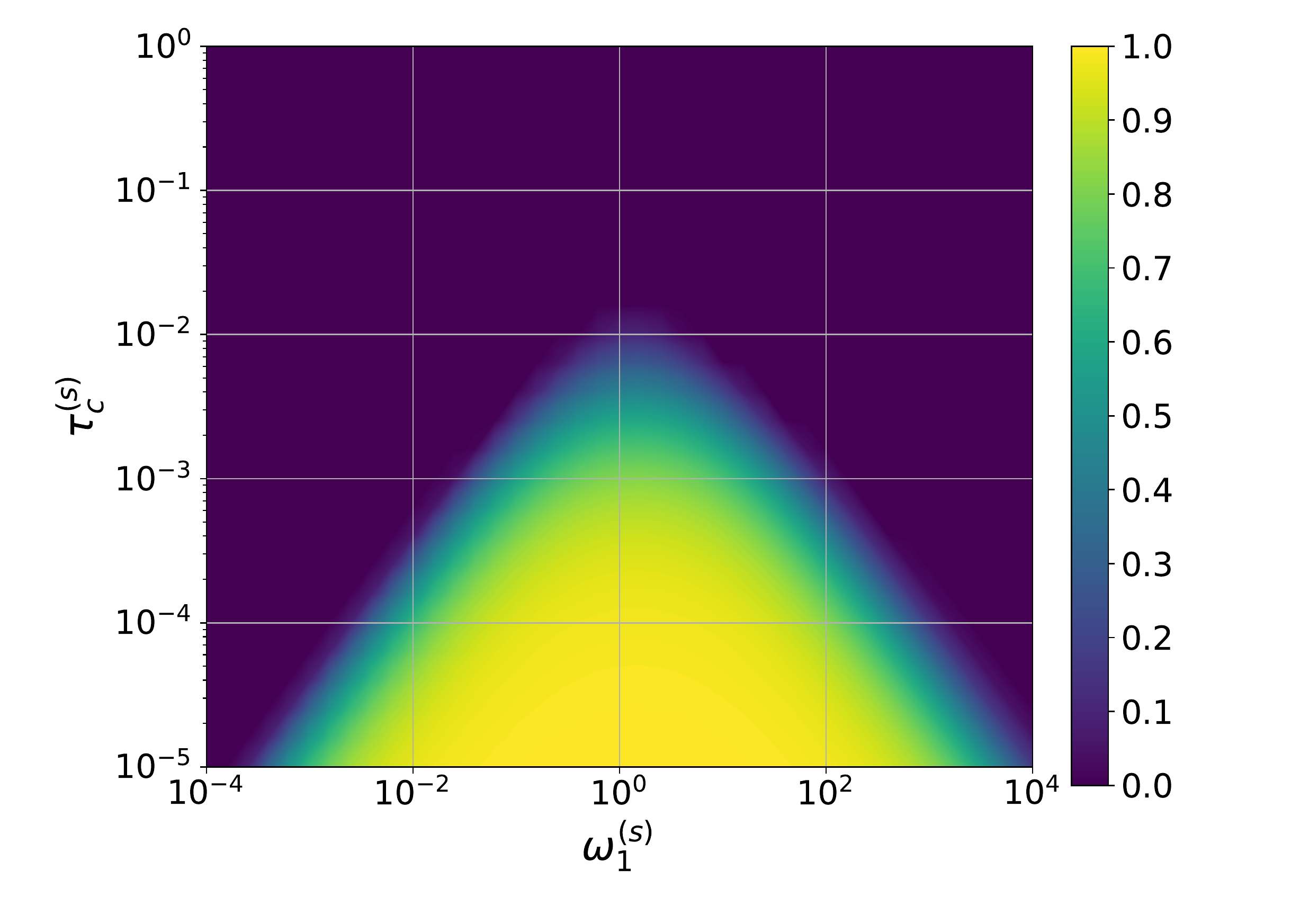}%\label{fig:DI_Concurrence}}\\
%\hspace{\fill}
%\subfloat[Efficiency of SWAP gate] % on $\omega_1$ and $\tau_c$]\
\raisebox{3cm}{(c)}\hspace{-1mm}
\includegraphics[width=0.4\textwidth, keepaspectratio]{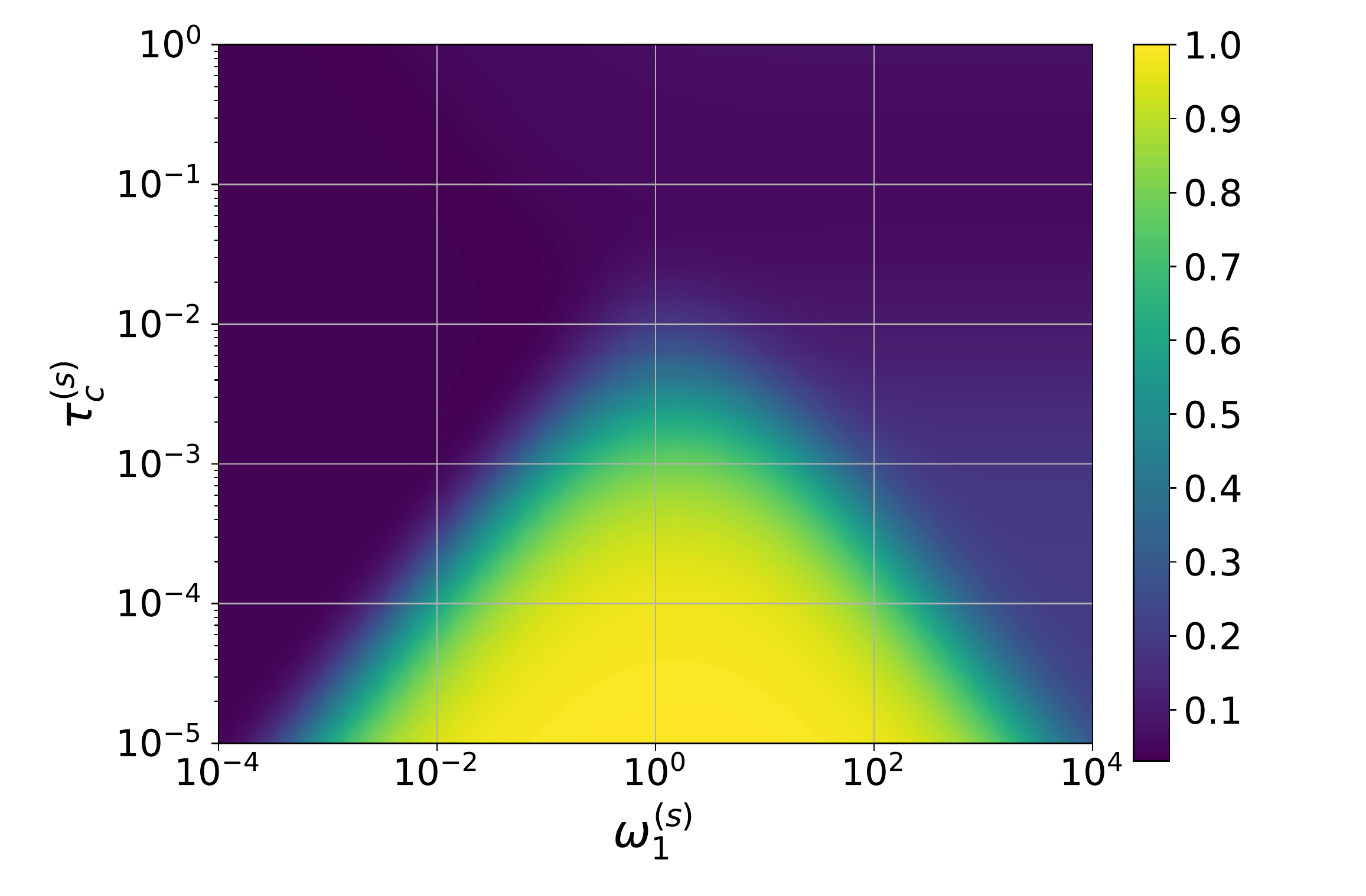}%\label{fig:DI_Efficiency}}
%\hspace{\fill}
%\subfloat[Fidelity of $|\psi_f\rangle$]% on $\omega_1$ and $\tau_c$
\raisebox{3cm}{(d)}\hspace{-1mm}
\includegraphics[width=0.4\textwidth, keepaspectratio]{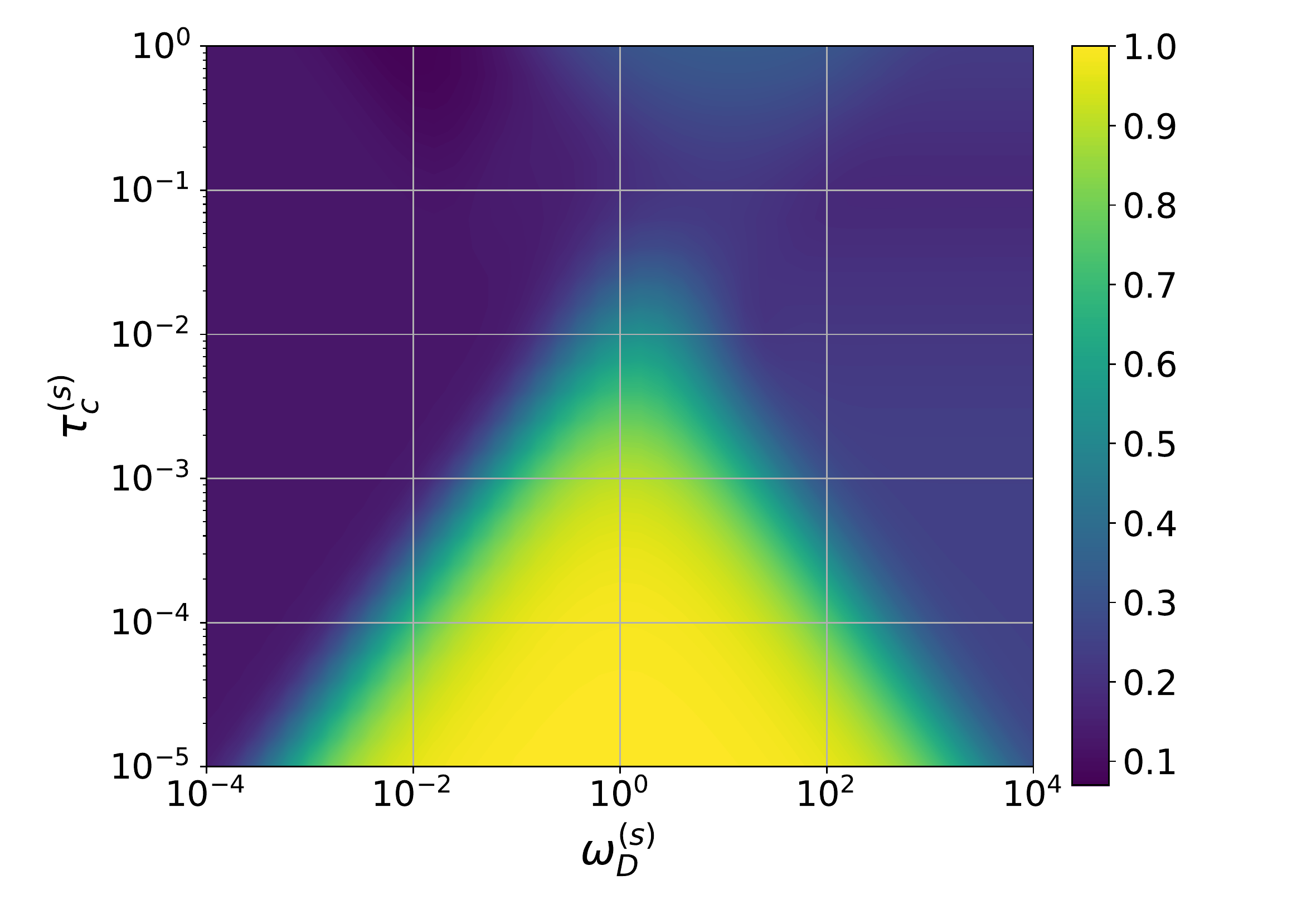}%\label{fig:DI_WD_Fidelity}}

\caption{Fidelity of expected state and efficiency as a function of drive power ($\omega_1$), strength of
dipolar coupling ($\omega_D = 2\pi J$) and characteristic time ($\tau_c$) of the bath (scaled to
system-environment coupling strength $\omega_{SE} $).  (a) Fidelity of $|\psi_f\rangle$, (b) Concurrence
between 2nd \& 3rd spin, (c) Efficiency of SWAP gate, (d) Fidelity of $|\psi_f\rangle$.  These simulated
results are for Larmor frequencies of the spins as $\omega_0^1 = 2\pi \times 10^4$ kHz = $\omega_0^3$,
$\omega_0^2 = 2\pi \times 10^3$ kHz and strength of system-bath coupling $\omega_{SE} = 2\pi \times 10^2$
kHz and pulses are applied on first spin.  (a), (b) \&
(c) are for $\omega_D = 2\pi \times 150$ kHz and (d) is
for $\omega_1 = 2\pi \times 150$ kHz.} 

\label{fig:DI_SWAP_sim}
\end{figure}

Fig.\ref{fig:DI_SWAP_sim} shows we can get very high efficiency as well as fidelity of expected state (and
hence high entanglement transfer) when drive's power ($\omega_1$) and dipolar coupling strength ($\omega_D$)
are similar to system-environment coupling strength ($\omega_{SE}$) and for certain characteristic time of
bath ($\tau_c$), though there is decoherence due to environmental effects. But when the Larmor frequencies
of the concerned spins are far off, secular approximation gives different Hamiltonian.  Hence the above
pulse sequence does not work as SWAP gate. Note,if we use short lived pules, we can ignore the pulse
durations compare to the time delays in between the pulses,  both the cases SWAP gate operation time
requires same amount of time, $\frac{7}{2J}$.

\section{Discussion} 

Quantum gate operations, like its classical counterpart,  have specific optimal clock speed
\cite{Chanda_2020}.  They showed that to get maximum fidelity the frequency of Rabi oscillation
($\omega_1$ here) work as effective clock speed and it is identical to single- as well as multiple-qubit
gates.  This work and we have used FRQME, though other known decay terms can be incorporated in this
analysis also \cite{Keldysh_1965}, \cite{Muller_2017}. 

Two level system (TLS) under an external drive gives rise to Rabi oscillation in first order. In the second
order, there are two instances when the drive act ($t_1, t_2$ and $t_1 > t_2$) with them being secular pair.
It is assumed the environment evolves under fluctuation and dephases in between $t_1$ and $t_2$ which is
incorporated by taking ensemble average \cite{Chakrabarti_2018}.  Hence we get drive induced dissipation
(DID) as an additional contribution. Also this formalism predicts a complex susceptibility kind of
contribution, where real part provides decay/absorption and imaginary dispersive one provides shift.

For any gate operations, one generally uses sequence of pulse sequences with different amplitude, frequency
and phase.  Here SWAP gate is constructed as a sequence of narrow square pulses of fixed parameters. For
each of these pulses, one can construct a suitable superoperator $\hat{\hat{\Gamma}}$ in Liouville space
\cite{Chanda_2020}. Sequential application of respective $\hat{\hat{\Gamma}}$’s will lead to the final
density matrix. This approach is particularly suitable for numerical evaluation of the propagator.
Otherwise, one can use the Dyson time-series for the pulses to calculate the final density matrix.

Complexity of multiqubit system is far more than that of single qubit's.  It is known that the overall
operation time of a specific task in quantum computation on a multiqubit network is limited by the strength
of the qubit-qubit coupling (J), as time required for an arbitrary transitional-selective pulse is
\cite{steffen_2001}. To be adequately selective, a square pulse must have a duration $\tau_p$ which is
inversely proportional to J, i.e.  $\tau_p \gtrsim  \frac{1}{J}$. This in turn indicates that the drive
amplitude, $\omega_1$ must be less than or of the order of J ( to keep the flip angle constant). Therefore,
to achieve maximum fidelity on such a multiqubit system, one must satisfy the condition $\omega_1^{opt} \sim
\omega_{SE}  \leq J$. 

\section{Conclusion}

We have presented a scheme for implementing SWAP operation on a chain of dipolar coupled spin halves in
presence of environmental decoherence.  The pulse sequence, that implements SWAP operation,  depends on the
Larmor frequencies of concerned spin halves, on which the gate operates, and coarse-grained time scale,
$\Delta t$,i. e. let $\Delta \omega_0$ be the difference Larmor frequencies, if $\Delta \omega_0 \cdot
\Delta t < 1$, we use pulse sequence for identical qubits, otherwise we use sequence for non-identical
qubits  We got maximum fidelity of transfer when drive amplitude and dipolar coupling strength is of the
order of system-environment coupling strength, namely  $\omega_1 \sim \omega_{SE}$ and $\omega_D \sim
\omega_{SE}$.  One can use SWAP operation as a building block of transport problems, which is important in
quantum information processing.  Hence we get maximum fidelity of transport when drive amplitudes and
dipolar coupling strength is of the order of system-environment coupling strength.

\section*{Acknowledgments}

GD gratefully acknowledges Council of Scientific \& Industrial Research (CSIR), India, for a research
fellowship (File no: 09/921(0327)/2020-EMR-I).

\bibliographystyle{unsrt}
\bibliography{references}

\end{document}